\documentclass[12pt,headings=big,numbers=noenddot,DIV=14,a4paper]{scrartcl}%

\pdfoutput=1

\usepackage{amsmath,amssymb,amsfonts}
\usepackage[normal,font=small,labelfont=bf,labelsep=period]{caption}
\usepackage[pdftex]{color,graphicx}
\usepackage{subfigure}
\usepackage[english]{babel}
\usepackage[compress]{cite}
\addtokomafont{disposition}{\rmfamily\boldmath}

\usepackage[dvipsnames]{xcolor}
\definecolor{darkblue}{rgb}{0,0.2,0.6}
\definecolor{darkgreen}{rgb}{0,0.4,0}

\usepackage[linktoc=page,bookmarks=false,colorlinks=false,
linkbordercolor=RoyalBlue,citebordercolor=ForestGreen,urlbordercolor=CornflowerBlue]{hyperref}

\numberwithin{equation}{section}
\DeclareFontFamily{OT1}{pzc}{}
\DeclareFontShape{OT1}{pzc}{m}{it}{<-> s * [1.10] pzcmi7t}{}
\DeclareMathAlphabet{\mathpzc}{OT1}{pzc}{m}{it}
\title{\vskip 30pt
       \textcolor{black}{ElectroWeak theory after the first LHC phase}}
\author{\normalsize Riccardo
Barbieri}
\date{\normalsize{\it  Scuola Normale Superiore and INFN, Piazza dei Cavalieri 7, 56126 Pisa, Italy}}

\begin{document}
\begin{titlepage}

\maketitle
\thispagestyle{empty}

\begin{abstract}
\centerline{\bf Abstract}\medskip
\noindent
I summarize the status of the ElectroWeak Interactions after the first phase of the Large Hadron Collider and I give an outlook on its possible developments.
\end{abstract}

\end{titlepage}

\section{The outcome of the first LHC phase}
\label{sec1}

The outcome of the first LHC phase, ended a few months ago, can be effectively summarized as follows:
\begin{itemize}
\item The discovery of the/a Higgs boson\cite{Aad:2012tfa, Chatrchyan:2012ufa}: a very major fact, although not unexpected.
\item No new particle produced, nor any new phenomena observed: a definitely unexpected evolution, so far.
\end{itemize}
As a result of this outcome, the  pending question on the entire field is clear.  Is the discovery of the Higgs boson\cite{Englert:1964et, Higgs:1964pj} the coronation of the Standard Model (SM) or a first step on a road yet largely unexplored? The pros for the former option are evident. The newly found resonance at 125 GeV of mass may well complete the spectrum of the SM by adding the only expected physical scalar particle. On the other hand, the reasons in favour of the latter option appear at least equally important if one looks at the Lagrangian of the SM in its part that depends on the Higgs doublet field $h$. In a synthetic notation
\begin{equation}
\mathcal{L}_\phi = |D_\mu h|^2 + \mu^2 |h|^2 - \lambda |h|^4 -   h\Psi_i \lambda_{ij}\Psi_j + h.c.
\label{h_eq}
\end{equation}
where $i,j = 1,2,3$ are generation indices and $\Psi_i$ is the collection of all matter fields in the $i$-th generation. The quadratic term in $h$ carries with it the famous (or infamous) naturalness problem, to which we shall have to return. One does not know if there is any dynamics behind the quartic term, as it is the case, e.g., in the Anderson theory of superconductivity. Last but not least the Yukawa coupling term, with its free $\lambda_{ij}$ parameters, hides the flavour puzzle.

Within the limited space available, a few remarks on the flavour problem are useful to make. The origin of flavour  breaking is unknown. A possible interpretation of CP violation measurements is that no new scale associated with flavour breaking exists below $10^4\div 10^5$ TeV. This is not a necessity, however, nor it is the most interesting case, in my view. An underlying flavour symmetry, suitably broken, may limit possible deviations from the Cabibbo Kobayashi Maskawa (CKM) pattern of flavour physics in the quark sector, characteristic of the SM, even in  presence of new degrees of freedom at the TeV scale, carrying flavour indices (squarks, composite fermions, etc). Yet such deviations, at $20\div 30\%$ level, are compatible with current bounds and must be looked for, since their search  is  both competitive with and complementary to the current direct searches of such new degrees of freedom\footnote{See Ref. \cite{Charles:2013aka} for a recent summary of current data and future prospects}. Analogous considerations apply to the lepton sector as well, with $\mu \rightarrow e + \gamma$ as a paradigmatic example.

\section{About naturalness, once again}
\label{sec2}

The reason why the absence of deviations from the SM, at least in this first phase of the LHC, has come as a surprise is of course related to the naturalness problem of the Higgs boson mass. The paradigm of naturalness\cite{Wilson:1970ag,  Gildener:1976ai,   Gildener:1976ih, t'Hooft, Maiani}, which has oriented much of the activity in theoretical particle physics in the last thirty years or so, becomes then the central issue and may even be put into question.  

The traditional way to state the naturalness problem of the Fermi scale is in terms of the radiative corrections to the Higgs boson mass, cutoff at a scale $\Lambda$: in the SM
\begin{equation}
\delta m_h^2 \approx (125~GeV)^2 (\frac{\Lambda}{500~GeV})^2
\label{div}
\end{equation}
with the normalization of 125 GeV chosen to match the measured value by ATLAS and CMS. Values of $\Lambda$ higher than 1 TeV or so lead therefore to large corrections to the Higgs boson mass, perhaps too large to be tolerated. Although perfectly sensible in an effective field theory approach, this way of stating the problem may require some clarification\footnote{At the risk of being repetitive, after more than three decades of discussion on the issue.}. After all - one  says  sometimes - aren't we supposed to talk only of physical renormalized quantities, with all divergences suitably reabsorbed?  Not to mention the celebrated absence  at all of quadratic divergences, like the one in Eq. (\ref{div}),  in the dimensional regularization scheme. 

\begin{figure}[t]
\begin{center}
\includegraphics[width=0.60\textwidth]{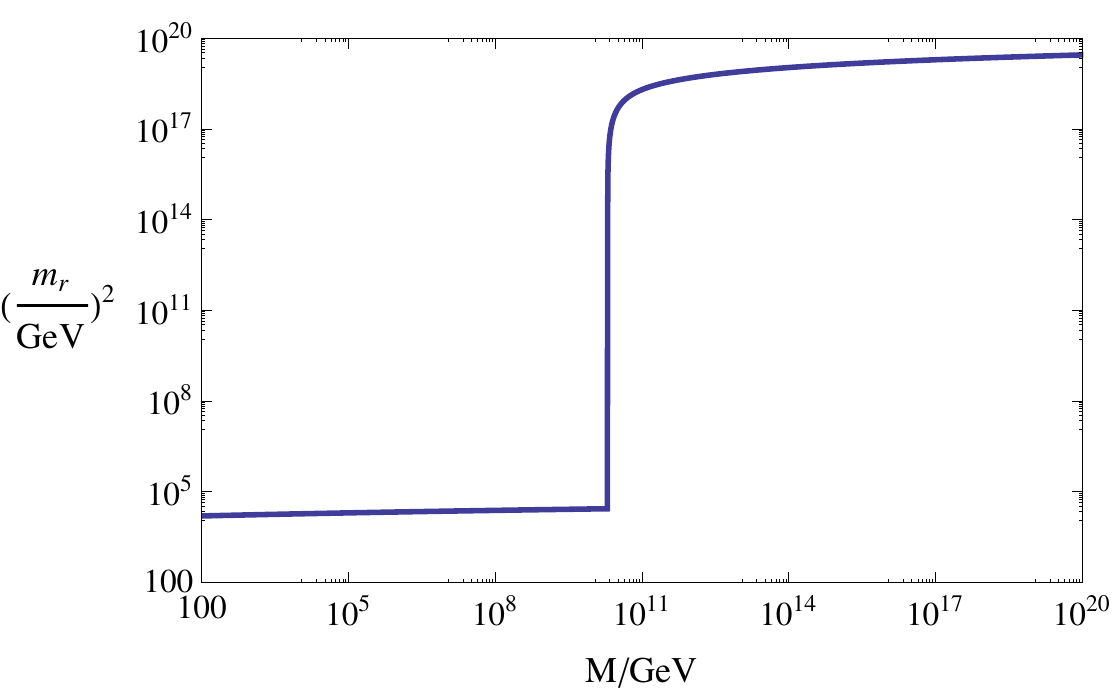}
\caption{\label{running} The running  Higgs mass squared versus the scale $M$ in the SM with the inclusion of a  particle of mass $M_H = 10^{10}~ GeV$ and a gauge invariant dimensionless coupling to the Higgs boson of strength $\lambda_H=1$. In general the jump involved at $M\approx M_H$ is of size $(\lambda_H M_H)^2/(16 \pi^2)$.}
\end{center}
\end{figure}
Indeed a neater way to state the naturalness problem is in terms of the renormalized running Higgs boson mass. (See, e.g., Ref. \cite{Barbieri:1996qp}). Fig. 1 shows the behavior of the Higgs mass squared versus the scale $M$ in the SM with the inclusion of a  particle of mass $M_H >> m_h$, coupled in a gauge invariant way to the Higgs boson through a dimensionless coupling  $\lambda_H$. The features in Fig. 1 are: i) a jump at the threshold $M \approx M_H$ of approximate size $(\lambda_H M_H)^2/(16 \pi^2)$; ii) a logarithmic behaviour of the running Higgs mass below and above the jump\footnote{The quadratic dependence  on $M_H$ of the jump is there irrespective of the nature of the particle coupled to the Higgs boson. For a $J=1/2$ particle the jump is on the negative side and for a $J=0$ state the dependence on $\lambda_H$ is linear. Some minor details of Fig. 1 depend on the precise definition of the running Higgs mass.}. The key point in Fig. 1 is that the "initial condition" on $m_r^2$ at some short distance scale, $M >> M_H$, has been chosen with great accuracy, of relative order $(m_h/M_H)^2$, in order to reproduce at $M = m_h$ the observed physical Higgs mass.
While this is technically possible, it is against the notion that the physics at the Fermi scale should not depend on  details of what happens at shorter distances, here at $1/M_H$. The quadratic divergence of the Higgs mass is not the problem {\it per se}, but the sign of the sensitivity of the Higgs mass to any threshold encountered at higher energy scales, like the one at $M_H$ in Fig. 1.

There are  at least three different ways to react to the naturalness or fine-tuning problem of the Fermi scale:
\begin{itemize}
\item Design a mechanism that protects the Higgs boson mass, no matter what the physics at shorter distances is.
\item Make assumptions on the short distance physics that may render it compatible with naturalness.
\item Accept the fine tuning, also based on the consideration of the cosmological constant  issue, which appears to present another very serious fine-tuning problem. 

\end{itemize}
%Let me briefly comment on each of them in the reverse order.

\subsection{A protected Higgs boson mass}

Supersymmetry\cite{Dimopoulos:1981zb, Dine:1981rt, Barbieri:1982eh, Chamseddine:1982jx, Hall:1983iz, AlvarezGaume:1983gj, Giudice:1998bp} or Higgs compositeness\cite{Kaplan:1983fs, Kaplan:1983sm, Agashe:2004rs, Giudice:2007fh} are  ways to protect the Higgs mass or the Fermi scale from being driven to whatever higher energy scale exists in nature. In both cases the searches for the new particles that are supposed to play the key role in the protection mechanism are well in place with several motivated configurations  being scrutinized (natural SUSY\cite{Dimopoulos:1995mi,
Cohen:1996vb, Barbieri:2009ev, Papucci:2011wy}, compressed spectra, R-parity violation in the quark sector\cite{Franceschini:2012za, Csaki:2013we}; composite top-like fermions with different branching ratios\cite{DeSimone:2012fs}). Broadly speaking,  the negative results of the first LHC phase so far set lower bounds on the masses of the relevant particles in the $500\div 1000$ GeV range.

\begin{figure}[t]
\begin{center}
\includegraphics[width=0.60\textwidth]{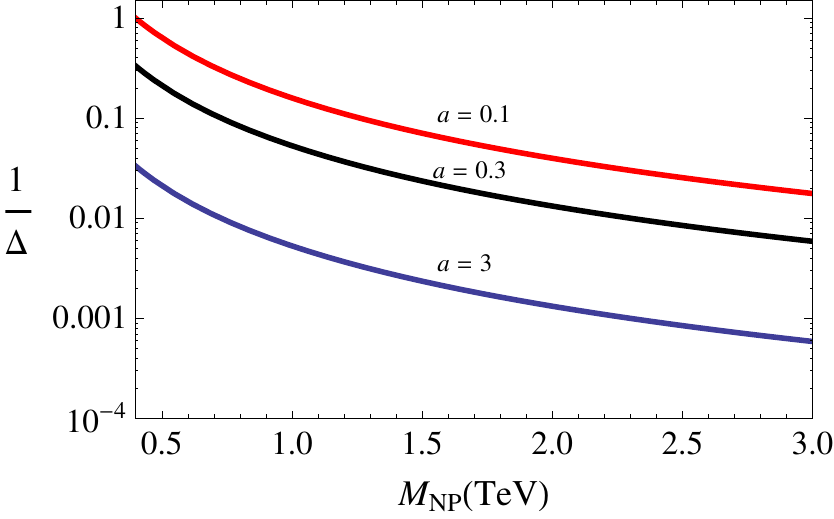}
\caption{\label{tuning} A semiquantitative measure of the fine tuning $1/\Delta$ in different models versus the  mass, $M_{NP}$, of the new particle that plays the main role in protecting the Higgs mass (e.g. a stop or a composite top-like fermion)}
\end{center}
\end{figure}

A global semiquantitative measure of the fine tuning implied by these bounds  is attempted in Fig. 2,  based on the equation for the correction to the Higgs mass 
\begin{equation}
\Delta  \equiv\frac{ \delta m_h^2}{ m_h^2}  \approx a \frac{M_{NP}^2}{ m_h^2},
\end{equation}
where $M_{NP}$ is the mass of the new particle that plays the main role in protecting the Higgs mass (e.g. a stop or a composite top-like fermion) and the dimensionless coefficient $a$ lumps all the dependence on a specific model (and may itself have a mild dependence on $M_{NP}$). $1/\Delta$ is  a measure of the amount of fine tuning that is necessary to accommodate a given $\delta m_h^2 > m_h^2$. The case $a=0.1$ (the one of least fine tuning in Fig. 2) would correspond to a model in which the cutoff $\Lambda$ in Eq. (\ref{div}) were replaced by $M_{NP}$. I am not aware of a model that achieves this without running into other problems, which incidentally  shows the significance of Eq. (\ref{div}) with the "naive" cutoff. The line at 
 $a=3$ is a fair representation of a Minimal Supersymmetric Standard Model (MSSM) with initial conditions at a large scale. The intermediate case $a=0.3$ can be reached in some Next to Minimal Supersymmetric Standard Model (NMSSM), to which I shall return. The bound $M_{NP} > 500\div 1000$ GeV implies in all cases some level of fine-tuning, which can even be stronger than $1\%$. At the LHC in its second phase one should be able to extend the search beyond  masses twice as large, thus exploring fine tunings at least four times stronger.

\subsection{Short-distance assumptions}

Since the fine tuning is a problem of high energy thresholds, a natural Higgs mass, instead of being automatically protected, may in principle arise by a precise selection  of every possible threshold felt by the Higgs boson itself\cite{Farina:2013mla}. For example a SU(5) Grand Unified Theory, non supersymmetric, with $M_H = M_{GUT}$ in Fig. 1, would not be allowed. A program like this has to digest at its start two difficulties. The first has to do with the existence of gravity and the related Planck mass. The second comes from the couplings in the SM that grow at high energy, certainly the hypercharge coupling $g_Y$ and perhaps some Yukawa coupling as well, like $\lambda_{top}$.  From these sources one would in fact expect corrections to the Higgs mass squared respectively proportional to $M_{Pl}^2$ and to $\Lambda_{Y, top}^2$, i.e. the scales at which  $g_Y$ or even $\lambda_{top}$ become non perturbative.

If one is willing to assume that these problems can find a solution\cite{Bardeen, Meissner:2006zh, Shaposhnikov:2007nj, Shaposhnikov:2009pv, Dubovsky:2013ira}, then one has still to constrain whatever BSM  physics to respect Higgs naturalness. The key is to keep under control the jump in Fig. 1 at $M\approx M_H$. This may happen either because $M_H$  is close enough to the Fermi scale or because the coupling to the Higgs boson, $\lambda_H$, is small enough, or a proper combination of both. Some of the problems in BSM physics, like Dark Matter or neutrino masses, may find solutions consistent with these requirements and may in some cases imply observable new physics at the TeV scale\cite{Farina:2013mla}.

\subsection{Accept the fine tuning}

In his well known review of the cosmological constant problem, at a time when the value of the cosmological constant was normally thought to be vanishing, Weinberg concludes by saying that "if it is only anthropic considerations that keep the effective cosmological constant within empirical limits, then this constant should be rather large, large enough to show up before long in astronomical observations"\cite{Weinberg:1988cp, Weinberg:1987dv}.  Given the observation of the accelerated expansion of the universe in 1998\cite{Riess:1998cb, Perlmutter:1998np} and later, which may be attributed to a non zero cosmological constant (more than $10^{120}$ times smaller than its "natural" value $\propto M_{Pl}^4$),  interest has arisen on the possibility  that also the weak scale may be fine tuned for similar "environmental" reasons\cite{Agrawal:1997gf}. This in turn almost inevitably leads to the view that contemplates the existence of an enormous number, say $N >>10^{120}$, of possible different  universes, or of different almost degenerate vacua of some fundamental theory, the so-called "multiverse".  

I do not address  the issue of the probability distribution of such universes\cite{Bousso:2007kq}, which will of course influence the actual value taken by fundamental parameters like the cosmological constant or the Fermi scale.
Rather it might be significant to note that the measured value of the Higgs boson mass may add  a new ingredient. If one extrapolates  the SM as it is up to energies close to $M_{Pl}$ (and assumes  no significant distortion from Planck-scale dynamics)\cite{Chetyrkin:2012rz, Bezrukov:2012sa, Degrassi:2012ry, Buttazzo:2013uya} the ElectroWeak vacuum is in a "near-critical" situation, not stable but meta-stable, i.e. sufficiently  long-lived to overcome the age of our universe. Rather then from anthropic considerations, one argues that such near-criticality might emerge from a probability density in multiverse space which favors critical points\cite{Buttazzo:2013uya}. Incidentally supersymmetry might play an important role here as well, although most likely not in the way discussed so far\cite{ArkaniHamed:2004fb, Giudice:2004tc, ArkaniHamed:2004yi, Giudice:2006sn, Hall:2009nd, Hall:2011jd}.

\section{One or more Higgs bosons? }
\label{sec3}

As mentioned in Section 2.1, the searches for the new particles that are supposed to play the key role in the protection mechanism of the Higgs boson mass are well in place and will be important in the next LHC phase. Among the other explorations that the experimental program will undertake, on general grounds I would rank high the search for possible extra scalars.
In contrast to this, here are the reasons that seem to speak in favour of a single Higgs boson, confronted with my reaction to them in each case:
\begin{itemize}
\item {\it Simplicity}

Why should the $J=0$ sector, with a single state, be so different from the $J=1/2$ and the $J=1$ sectors, both in terms of number of states and of irreducible representations of the gauge group?

\item {\it Electromagnetism unbroken}

Although it is true that with a single Higgs doublet electromagnetism is never broken,  the multi-doublet case only adds one phase ($SU(2)\times U(1)$ fully broken) to the two phases in the single doublet case ($SU(2)\times U(1)$ unbroken, $U(1)_{em}$ unbroken).

\item {\it The CKM picture automatically implemented}

As said in the first Section, there is no reason to be proud of the $\lambda_{ij}$ parameters.

\item  {\it A single tuning, in case}

No fine tuning is better, as, e.g., in the case of supersymmetry, which requires at least two Higgs doublets. Furthermore, naturalness implemented by short-distance assumptions allows  in principle  the presence of any number of scalars at the TeV scale\cite{Dubovsky:2013ira}. 

\end{itemize}
To orient  the search for extra scalars or to describe the corresponding results in useful terms is not easy. I think that one should proceed by suitable simplified models. As an example I  briefly discuss the case of the NMSSM, after recalling why  it is particularly motivated.

\subsection{The NMSSM case}

\begin{figure}[t]
\begin{center}
\includegraphics[width=0.60\textwidth]{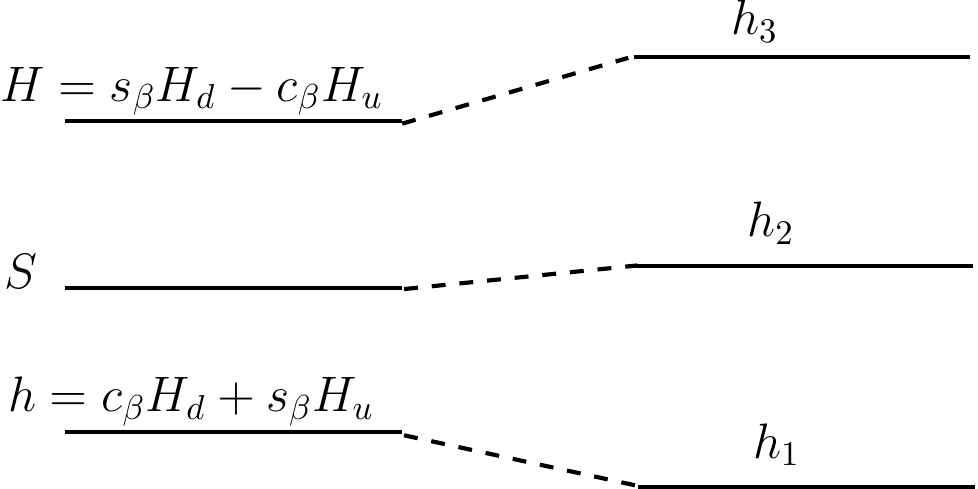}
\caption{\label{spectrum}Right side: The spectrum of the $J^P = 0^+$ neutral states of the NMSSM.  Left side: states with definite electroweak quantum numbers before mixing. The ordering of the levels is arbitrarily chosen.}
\end{center}
\end{figure}

The NMSSM is the simplest extension of the MSSM, where the inclusion of a singlet "chiral" multiplet $S$ allows to write the Yukawa coupling $\lambda_S S H_u H_d$ in a way consistent with supersymmetry\footnote{I use $\lambda_S$ not to confuse it with the quartic coupling of the SM. $H_u$ and $H_d$ are the Higgs doublet multiplets with a Yukawa coupling to the up and to the down quarks respectively.}\cite{Fayet:1974pd, Ellwanger:2009dp}.
There are two independent reasons to consider the NMSSM, both related to naturalness but independent from each other. One is well known and has to do with the expression for the upper bound on a scalar mass in the spectrum of the $J=0$ particles
\begin{equation}
m_h^2 =  m_Z^2 \cos^2{2\beta} + \lambda_S^2 v^2 \sin^2{2\beta} + \Delta_t^2,
\end{equation}
with the first and the third term in the r.h.s being the tree-level and the loop-correction  in the MSSM. The presence of the extra term proportional to $\lambda_S^2$ allows to accommodate a scalar with SM-like properties at 125 GeV, as the one observed, without having to resort to a large $\Delta_t$, i.e. heavy and/or strongly mixed stops, not liked by naturalness. The second reason comes from the fact that the Higgs vev squared at tree level is proportional to $ 4 M_{NP}^2/ (g^2 + g^{\prime 2})$ in the MSSM, whereas in the NMSSM, at moderate $\tan{\beta}$ and $\lambda_S$ close to unity, this same expression gets replaced by $M_{NP}^2/\lambda_S^2$.   This is why the NMSSM, relative to the MSSM,  can accommodate heavier s-particles (stops, gluinos) of typical mass $M_{NP}$ at the same level of fine tuning\cite{Harnik:2003rs, Barbieri:2006bg, Perelstein:2012qg, Gherghetta:2012gb, Lu:2013cta}. In turn this makes it not unconceivable that the extra scalars of the NMSSM be the lightest new particles around.

\begin{figure}
\begin{center}
\includegraphics[width=0.48\textwidth]{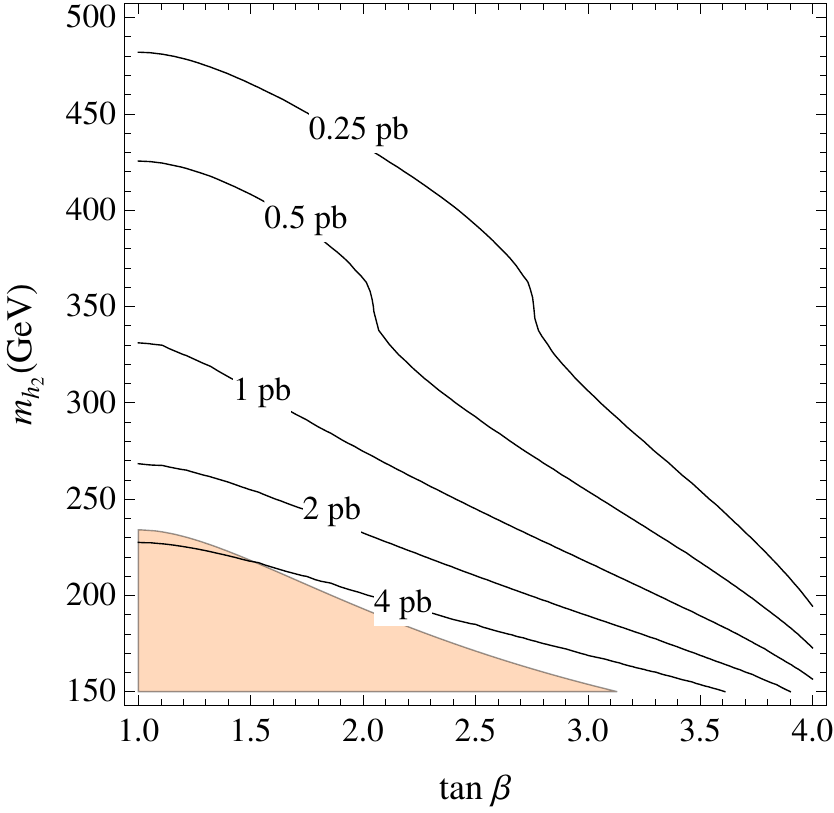}\hfill
\includegraphics[width=0.48\textwidth]{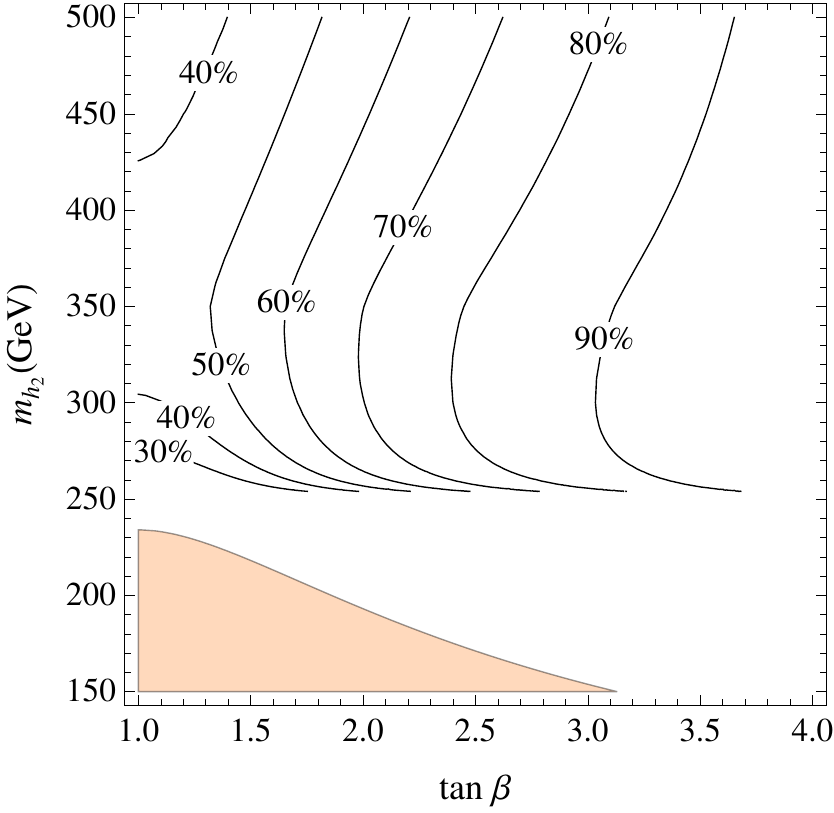}
\caption{\label{fig:mAdecoupled-1}\small $H$ decoupled, $\lambda=0.8$. Left:  Isolines of the gluon fusion cross section $\sigma(gg\rightarrow h_2)$ at LHC14. Right: Isolines of $BR(h_2\rightarrow h_1 h_1)$. The colored region is excluded at 95\%C.L. by current experimental data for the signal strengths of $h_1 = h_{\rm LHC}$.}
\end{center}
\end{figure}

The presence of extra scalars can show up in two distinct ways: i) by direct production; ii) indirectly through 
the couplings of the already observed resonance  at 125 GeV, due to mixings between states of different electroweak charges. Assuming negligible CP violation in the Higgs sector,
the mixing scheme in the $J^P = 0^+$ neutral states of the NMSSM is illustrated in Fig. 3. On the left side are the states with definite electroweak quantum numbers: two doublets and one singlet. The two doublets are further defined by their vev: 0 for $H$ and $v$ for $h$, which makes its couplings to the fermions or to the gauge bosons identical to the ones of the Higgs boson in the SM. This states and the singlet $S$ are  in general mixed to form the mass eigenstates, $h_1, h_2, h_3$, as shown in the right side of Fig. 3. 

To organize and to understand the impact of the search of the extra scalar states, I think that it is best to distinguish four different limiting situations in which $h_1$ is identified with the state already found at LHC, $h_{\rm LHC}$, and can be the lightest or the next-to-lightest state:

\begin{itemize}
\item Singlet-decoupled, $h_3  < h_{\rm LHC} < h_2 (\approx S)$
\item Singlet-decoupled, $h_{\rm LHC} < h_3  < h_2 (\approx S)$
\item $H$-decoupled, $h_2  < h_{\rm LHC} < h_3 (\approx H)$
\item $H$-decoupled, $h_{\rm LHC} < h_2  < h_3 (\approx H)$
\end{itemize}

In each case the production cross sections and the branching ratios are characterized in term of few parameters\cite{Barbieri:2013hxa, Barbieri:2013nka}. At the same time  it is easy to see in this simplified space the impact of the measurements of the signal strengths of $h_{\rm LHC}$. 
Such measurements, both present and foreseen, are powerful in the Singlet-decoupled case in which the two doublets $h$ and $H$ are mixed together. They are less important  when the doublet $h$ is mixed to the singlet $S$ ($H$-decoupled). In this case the decay $h_2\rightarrow   h_{\text{LHC}} h_{\text{LHC}}$ might be an important discovery channel, with production cross section and branching ratio illustrated in Fig. 4 for the second LHC phase\cite{Barbieri:2013hxa}. Both in the Singlet-decoupled or in the $H$-decoupled case, one can easily convince oneself that the existence of extra Higgs states is not affecting the  ElectroWeak Precision Tests currently available at a level that can compete with the direct searches or with the measurements of the signal strengths of $h_{\rm LHC}$\cite{Barbieri:2013nka}.

\section{Summary}
\label{sec4}

At least among theorists, the prevailing attitude in the last three or four decades has been to consider the SM as a low energy effective description of a more fundamental theory at shorter distances. A part from motivations of general order (the Wilsonian approach to field theory, etc.) the reasons behind this view have to do with the problems related to Eq. (\ref{h_eq}) and recalled in Section 1. None of these problems introduces any physical inconsistency but the reasons for the discontent about them are real. The first thorough experimental exploration of the Fermi scale by the LHC was/is supposed to clarify at least some of these problems: a presumption based on naturalness.
This has not happened in the first LHC phase so far, in spite of the very major discovery of the/a Higgs boson. Not surprisingly, therefore, the paradigm of naturalness becomes a central issue and may even be put into question.
If properly intended, however,  it cannot be dismissed on the basis that it is unsound. 

Three different possible reactions to this situation have been recalled in Section 2:
\begin{itemize}
\item {\it Insist on natural theories, whatever the physics at short distances is.}

There is no objective way to tell which amount of fine tuning is tolerable. Fine tunings exist in nature. On the other hand it is a fact that in  other cases where some appropriate physics enters to cure a naturalness problem (the positron for the classical electron self energy, the $\rho$ meson for the $\pi^+-\pi^0$ mass difference and the charm quark for the neutral kaon mass difference) no fine tuning is involved at all. All together, I think that the case for supersymmetry or Higgs compositeness is still open and might deserve positive surprises at LHC in its second phase. To this end the search for deviations from the CKM picture of flavour and CP violation, of paramount importance {\it per se}, could also play an indirect role.

\item {\it Select (and make assumptions about) the short distance physics that can be compatible with naturalness.}

As I have tried to make clear, the SM in isolation would be a perfectly natural theory. Why then the big resistance to take this option seriously? Because of gravity, to the least, and because of the  couplings growing with energy in the SM alone. If one can get around these potential problems, one may  interpret the current situation as providing a criterium of strong selection of every possible new physics at high energies, like $M_H$ in Section 2. The key is to keep moderate the jump in Fig. 1. Since low $M_H$'s are preferred, this also motivates exploring the TeV scale as best as one can.

\item {\it Accept the fine tuning.}

To the extent that good physical theories are recognized by their ability to make predictions, the statement by Weinberg, recalled  in Section 2.3, about the cosmological constant  may contain a message relevant to the case of the Fermi scale too. The peculiar position in the $(m_h, m_t)$ plane of the observed universe, with the SM extrapolated as it is up to $M_{Pl}$,\cite{Cabibbo:1979ay} might  in fact also invite another  "environmental" interpretation, logically different  from the anthropic one: a preference in the "multiverse" for "critical" points. How is all to be seen. I admit to be frightened by the apparent difficulty, at least so far,   to see some unambiguous experimental test of the environmental selection of the Fermi scale.

\end{itemize}

Some reader may be disturbed by the insistence on the naturalness issue in this brief review of the status of the theory of the ElectroWeak interactions. I think on the contrary that this is justified by the role, if not by anything else, that naturalness has played so far. Rightly so, I believe. Which does not mean that alternative roads should not be pursued, if one can. As a relevant example, I have indicated on general grounds the  interest of looking for extra scalar states. They may or  may not be a manifestation of natural theories, like the MSSM or the NMSSM. To organize the search of these extra states and to describe the corresponding results it will be best to proceed by suitable simplified models, in an analogous way to what has been and is being done in the search for other particles. 

\subsubsection*{Acknowledgments}
It is a pleasure to thank the Organizers of the LHC Nobel Symposium, in particular Prof. Tord Ekelof. This work is supported in part by the European Programme ``Unification in the LHC Era",  contract PITN-GA-2009-237920 (UNI\-LHC) and by MIUR under the contract 2010YJ2NYW-010.
%
%\bibliographystyle{My}
%\small
%\bibliography{Higgses_biblio}

\end{document}